# Influence of $BaTiO_3$ nanoparticles on the anisotropy of the dielectric properties of nematic liquid crystal 5CB

J. M. Gudenko[1], ORCID: https://orcid.org/0009-0005-5938-5480, O. S. Pylypchuk[1], ORCID: https://orcid.org/0000-0003-0136-0799, V. V. Vainberg[1*],ORCID: https://orcid.org/0000-0002-9840-8033, D. O. Stetsenko[1], ORCID: https://orcid.org/0009-0002-9141-0939, I. A. Gvozdovskyy[1], ORCID: https://orcid.org/0000-0002-4531-9942, S. E. Ivanchenko[2], ORCID: https://orcid.org/0000-0002-9859-1835, E. A. Eliseev[2], ORCID: https://orcid.org/0000-0001-8124-8857, V. N. Poroshin[1†],ORCID: https://orcid.org/0000-0001-8217-3949, and A. N. Morozovska[1‡], ORCID: https://orcid.org/0000-0002-8505-458X.

[1] *Institute of Physics of the National Academy of Sciences of Ukraine,46, Nauky Ave., 03028 Kyiv, Ukraine*

[2] *Frantsevich Institute of Materials Science, the National Academy of Sciences of Ukraine, 3, Omeliana Pritsaka, 03142 Kyiv, Ukraine*

## Abstract

This work is devoted to the mechanisms of dielectric response and electric conductivity of suspensions consisting of the nematic liquid crystal 5CB with different concentrations (from 0 to 10 wt.%) of the ferroelectric $BaTiO_3$ nanoparticles with an average size of 24 nm. We revealed that the incorporation of nanoparticles influences significantly the dielectric permittivity magnitude and anisotropy, as well as dielectric losses of the suspension. A pronounced temperature dependence of the anisotropic dielectric permittivity of the suspensions was found at lower temperatures corresponding to the mesophase state; but it is also present at higher temperatures corresponding to the isotropic phase. The dependence of the mesophase – isotropic phase transition temperature on the concentration of $BaTiO_3$ nanoparticles appeared nonmonotonic. With increasing temperature, both the capacitance and the electrical resistance of the pure liquid crystal increase, as well as they increase in the suspensions with small concentration of $BaTiO_3$ nanoparticles. Due to space charge accumulation in the shells of nanoparticles, larger concentrations of $BaTiO_3$ nanoparticles influence strongly the ionic transport by promoting the formation of ionic-electronic screening. This effect modifies the dielectric properties and conduction mechanisms of the suspension, leading to the nonmonotonic dependence of the mesophase – isotropic phase transition temperature versus the nanoparticle concentration.

[*] corresponding author, e-mail:viktor.vainberg@gmail.com

[†] corresponding author, e-mail: vnporoshin17@gmail.com

[‡] corresponding author, e-mail:anna.n.morozovska@gmail.com

## 1. *Introduction*

Liquid crystal (LC) suspension containing ferroelectric nanoparticles (NPs) is an extremely interesting and promising object for fundamental research on the influence of particle concentration and interparticle interactions on the polar and dielectric properties of the suspension [1, 2, 3]. Such systems allow us to study the long-range dipole-dipole interactions [4, 5], the behavior of polarized nanoparticles and ionic transport in soft anisotropic media [6, 7].

Of particular interest is the electrostatic interaction of the liquid crystal molecules with ferroelectric nanoparticles in the suspensions, which determine the anisotropy of the dielectric properties [8, 9, 10]. The joint action of the temperature-induced phase transitions in nematic liquid crystals and anisotropic dipole-dipole interactions with suspended ferroelectric nanoparticles opens wide opportunities for the creation of new soft materials with controllable glass-type dynamics of dielectric properties [11] and electro-optical features [12]. Therefore, these suspensions are important from the point of view of advanced applications: they can be used as basic elements for creation of modern non-volatile memory cells, and energy storage devices, as well as for optical and optoelectronic applications [10-12].

Also, of considerable fundamental and practical interest are the studies of the electrophysical properties of ferroelectric nanoparticles with sizes from 5 to 50 nm, which are suspended in a liquid dielectric medium [13, 14]. Important, that ferroelectric $BaTiO_3$ nanoparticles covered by heptane and oleic acid acquire a core-shell type structure [15, 16, 17] and, being suspended in a liquid dielectric medium, reveal a "giant" spontaneous polarization (about 1.3 C/m$^2$ at the room temperature). The core-shell nanoparticles with the giant spontaneous polarization can significantly influence the phase transitions dynamics in the nematic liquid crystals [4-17] and enhance the electrocaloric response of colloids [18]. It was subsequently shown that the lattice constant mismatch, which causes epitaxial and/or core-shell strains, is responsible for the enhancement of spontaneous polarization and tetragonality in the core-shell $BaTiO_3$ nanoparticles of the spherical [19, 20, 21] or cylindrical [22, 23] shape.

In this work we study the features of the dielectric response and electric conductivity of suspensions consisting of the nematic liquid crystal 5CB (4-pentyl-4'-cyanobiphenyl) with different concentrations (from 0 to 10 wt.%) of the $BaTiO_3$ nanoparticles. We have shown that a large concentration of the $BaTiO_3$ nanoparticles influences strongly the ionic transport by promoting the formation of ionic-electronic screening due to the space charge accumulation in the shells of nanoparticles. This effect modifies the dielectric permittivity, its anisotropy and dielectric losses, as well as determines the conduction mechanisms of the suspension, leading to the nonmonotonic dependence of the meso-isotropic phase transition temperature on the concentration of the $BaTiO_3$ nanoparticles.

## 2. *Sample preparation and experimental methods*

To prepare the samples, a nematic LC, abbreviated 5CB, was used. The technique of cells with LC in full detail was described in detail in Ref. [6]. A barium titanate ($BaTiO_3$) nanopowder consists of nanoparticles having a quasi-spherical shape and an average particle size of 24 nm and a relatively small dispersion of sizes (less than ±(10 – 15) nm). The thickness of the studied cells was 21 ± 0.5 μm. A series of samples with different concentrations of nanoparticles, namely 0.5, 1, 2, 3, 5 and 10 wt. % $BaTiO_3$ have been prepared. Note that the bulk $BaTiO_3$ has a large anisotropic relative dielectric permittivity at the room temperature: about 120 – 170 in the polar direction and about 2000 – 4000 in the transverse directions. According to X-ray diffraction results and theoretical calculation performed earlier for the same $BaTiO_3$ nanoparticles [18, 21], the large nanoparticles (size ~ 15 – 40 nm at the room temperature 20ºC) are in the single-domain ferroelectric state; and the smaller nanoparticles (size ~ 9 – 14 nm at 20ºC) are in the paraelectric phase due to the size-induced phase transition to the nonpolar phase. The single-domain particles behave as individual dipoles in the liquid crystal environment, partially or completely screened by the shell of ionic-electronic charge, while the paraelectric nanoparticles may become a source of the ionic-electronic charge under the temperature increase above the temperature of the size-induced phase transition [24].

**Figure 1** shows schematically the experimental setup. The sample is located inside a thermostabilized chamber, which allows measurements to be made at a controlled temperature that can be varied over a wide range, which is important for studying the nematic-isotropic phase transition and the temperature dependence of dielectric permittivity and losses. The LCR meter LCX200 ROHDE & SCHWARZ was used to measure the capacitance and electrical resistance as a function of the frequency, applied voltage, and temperature.

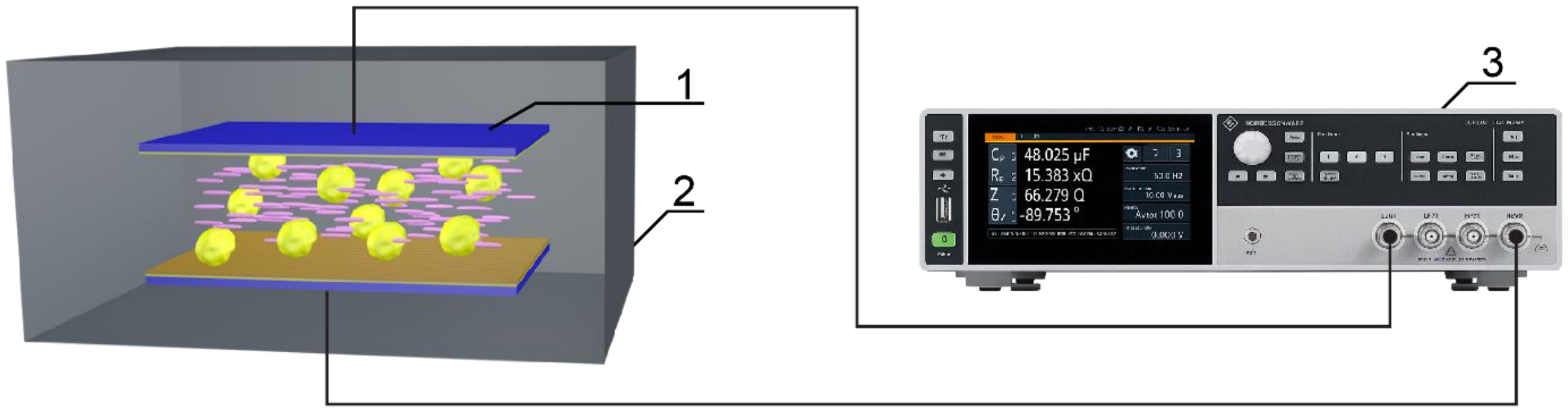

**Figure 1.** Schematic of experimental setup. **1)** Measurement sample, **2)** temperature-controlled chamber, **3)** LCR meter.

**Figure 2** shows schematically the orientation of 5CB liquid crystal molecules with ferroelectric $BaTiO_3$ nanoparticles in the cell without and under large electric voltage bias perpendicular the cell plane that creates the strong electric field capable to orient the 5CB molecules. Without the orienting electric field, the 5CB molecules are oriented mainly parallel to the substrate surface, because such orientation is dictated by the preparation conditions of the liquid crystal cell. The yellow spheres show the $BaTiO_3$ nanoparticles dispersed in the 5CB liquid crystal. Under the electric field exceeding the critical value, the Fréedericksz transition occurs: the director of the liquid crystal changes its direction, and the 5CB molecules align along the electric field normal to the electrode surfaces of the cell.

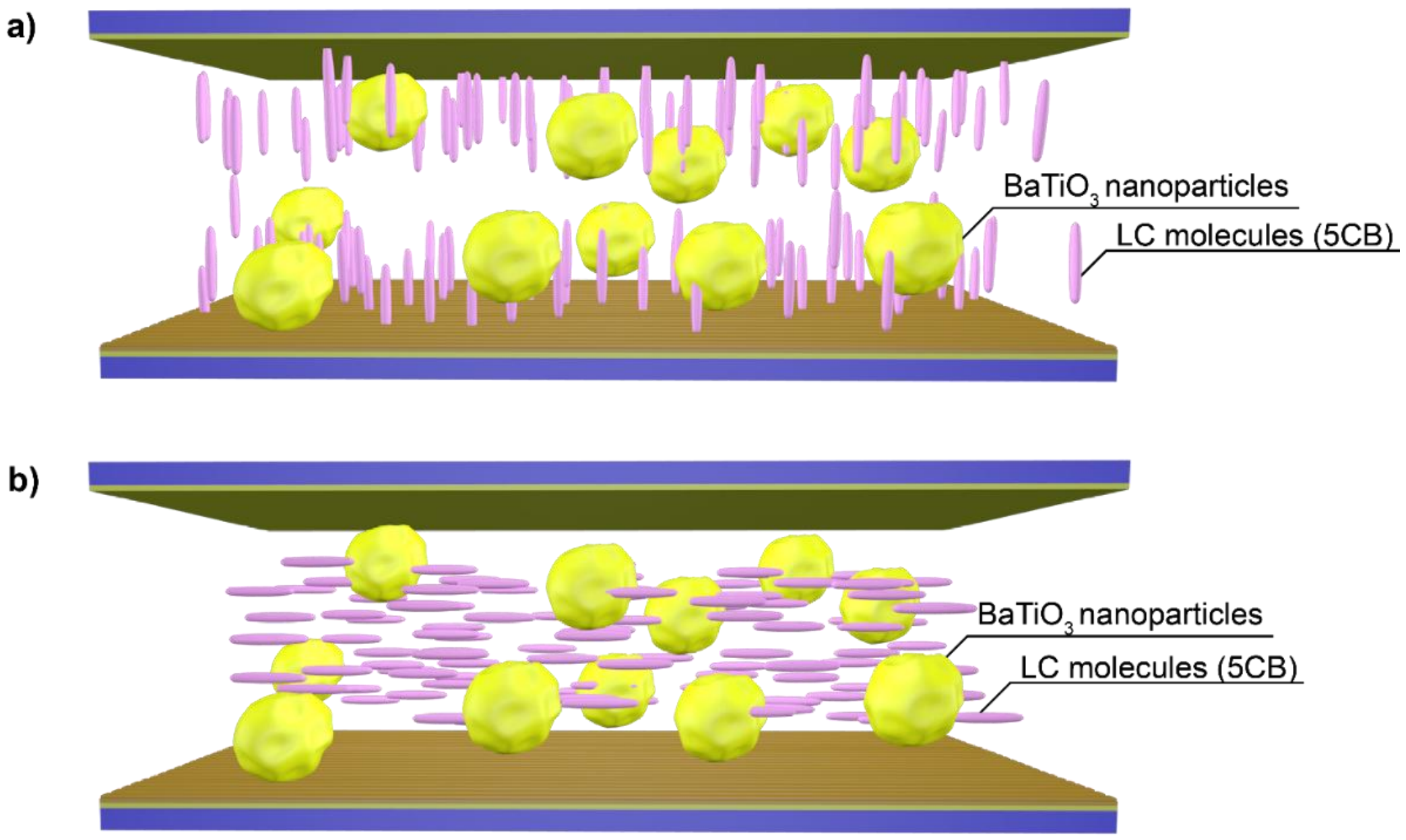

**Figure 2.** Schematic orientation of the 5CB liquid crystal molecules and $BaTiO_3$ ferroelectric nanoparticles in the cell under large bias voltage (**a**) and without bias (**b**). Golden spheres are $BaTiO_3$ nanoparticles and blue ellipses are 5CB liquid crystal molecules.

### 3. *Experimental results and discussion*

**Figure 3** presents the frequency dependences of the effective dielectric permittivity and dielectric loss tangent for the 5CB liquid crystal with different concentrations of the $BaTiO_3$ nanoparticles. The measurements were performed at the temperature of 20°C and measurement signal amplitude of 0.25 V. Large values of the dielectric permittivity are observed in the low-frequency region. This is due to the contribution of ions and interfacial polarization. With increasing frequency, the dielectric permittivity decreases, since slow polarization mechanisms no longer have time to follow the changing electric field. The influence of the nanoparticle concentration on the dielectric response is particularly interesting. For concentrations of 0.5%, 1%, 2% and 3%, the dielectric permittivity is higher than that of a pure liquid crystal in a wide frequency range.

This trend can be explained by two main mechanisms. Firstly, the $BaTiO_3$ nanoparticles in the ferroelectric state, and especially those in the paraelectric phase, have a colossal dielectric permittivity that is hundreds of times larger than that of 5CB. Therefore, even a small concentration of the $BaTiO_3$ nanoparticles enhances the effective polarization response of the suspension. Secondly, the single-domain nanoparticles behave as electric dipoles. As a result, the dipole-dipole interactions arise between the nanoparticles and the liquid crystal molecules, which improve the local ordering of molecules and enhance the dielectric anisotropy of the system.

However, the situation changes at large concentrations of the $BaTiO_3$ nanoparticles, namely at 5 % and 10 %, because the effective dielectric permittivity decreases for these concentrations and becomes even lower than that of a pure liquid crystal in some frequency ranges, which may indicate a change in the mechanism of the system dielectric response. One of the possible reasons may be the aggregation of nanoparticles and formation of clusters, as well as a decrease in the distance between nanoparticles, which can form partially connected structures or charge flow channels. In this case, the system begins to behave not as a capacitor, but as a heterogeneous resistive-capacitive medium and some part of the external field energy is spent not on its polarization, but on the charge transfer through conductive channels. In addition, at high concentrations, the role of ionic-electronic screening charges at the surface of nanoparticles increases, which compensate (at least partially) the spontaneous polarization of the $BaTiO_3$ nanoparticles and reduce the effective dipole moment of nanoparticles. The dielectric loss tangent retains a pronounced maximum for all samples. The position and height of the loss tangent maximum changes with increase in nanoparticle concentration indicating the change in the mechanisms of charge transfer and polarization response.

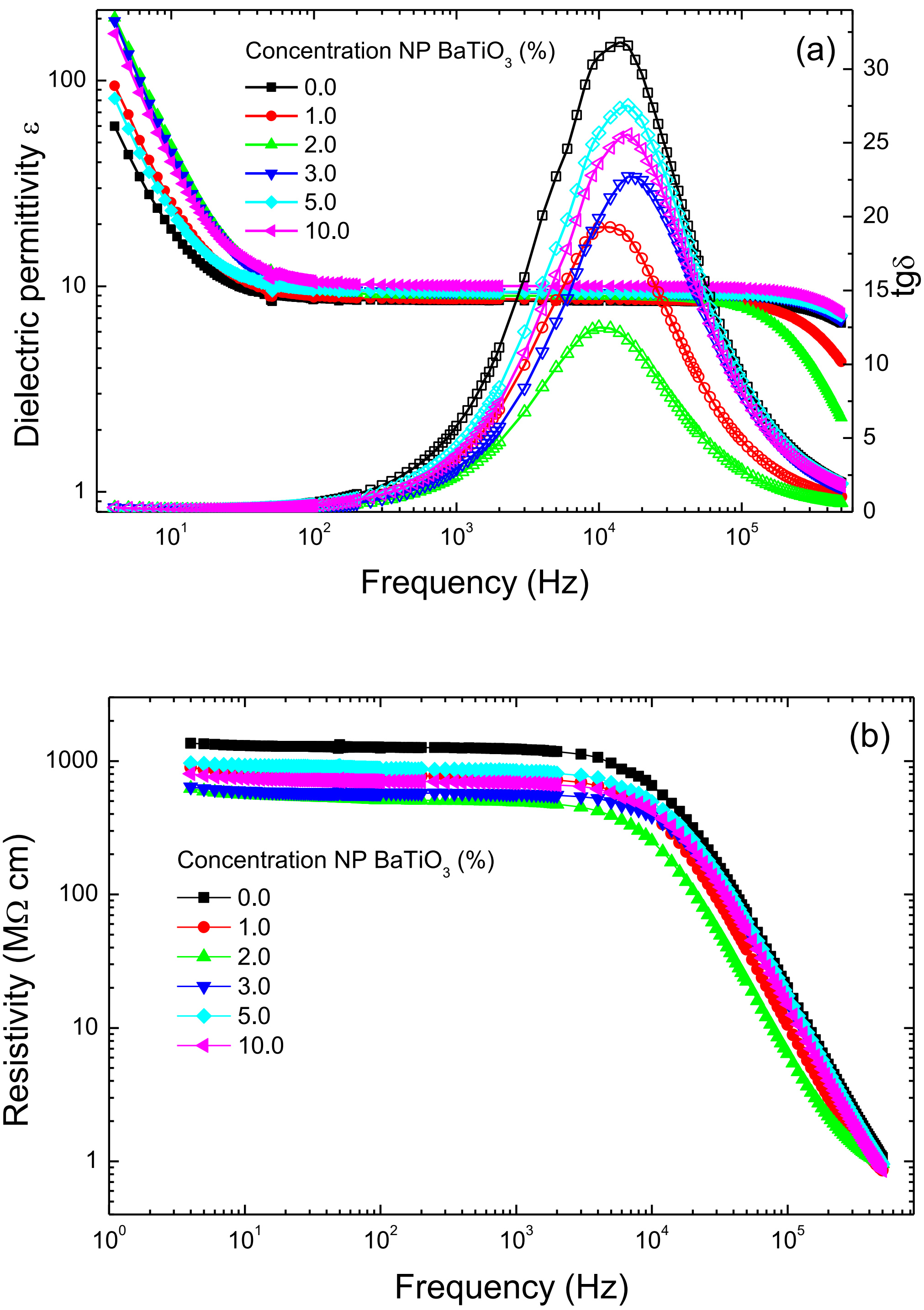

**Figure 3.** Frequency dependences of the effective dielectric permittivity and dielectric loss tangent **(a)** and resistivity **(b)** of 5CB liquid crystal with different concentrations of the $BaTiO_3$ nanoparticles measured at 20°C. The voltage level is 0.25 V.

**Figure 4(a)** shows the dependences of the dielectric permittivity of liquid crystal suspensions with the $BaTiO_3$ nanoparticles on the electric field strength created by the external voltage bias perpendicular to the cell plane. The dielectric permittivity remains almost constant in the range of small electric fields. When the field strength exceeds approximately 1.8 – 2.2 kV/cm, a sharp increase in permittivity is observed due to the Fréedericksz transition, because the electric field is sufficient to reorient the liquid crystal molecules [8-10]. The presence of $BaTiO_3$ nanoparticles significantly affects this process. For concentrations of nanoparticles 1 – 5%, the permittivity, measured after the Fréedericksz transition, becomes greater than that of a pure liquid crystal. Important that the magnitude of permittivity increase and the position of the Fréedericksz transition (electric field region) depends on the concentration of the $BaTiO_3$ nanoparticles, because the nanoparticles affect the threshold voltage of director reorientation and the electro-optical behavior of the system. However, for the concentration of 10%, the permittivity increases weaker and slower.

**Figure 4(b)** shows the dependences of the resistivity of liquid crystal suspensions with the $BaTiO_3$ nanoparticles on the electric field strength on swiping the applied voltage up and down. For all samples the nonmonotonic behavior is observed, since the reorientation of the liquid crystal molecules can change the mobility of ions, the local spatial distribution of the electric field, and the conditions of charge transfer through the cell. It is especially important that the addition of the $BaTiO_3$ nanoparticles significantly changes the resistance magnitude. For some concentrations, the resistance becomes significantly lower. One possible explanation may consist in the formation of local charge transfer channels between the nanoparticles. Also note that the branches of these curves, corresponding to increasing and decreasing voltage, strongly differ from each other by magnitude, which is caused by a slow enough relaxation process, and is interesting from the point of view of resistive switching effect.

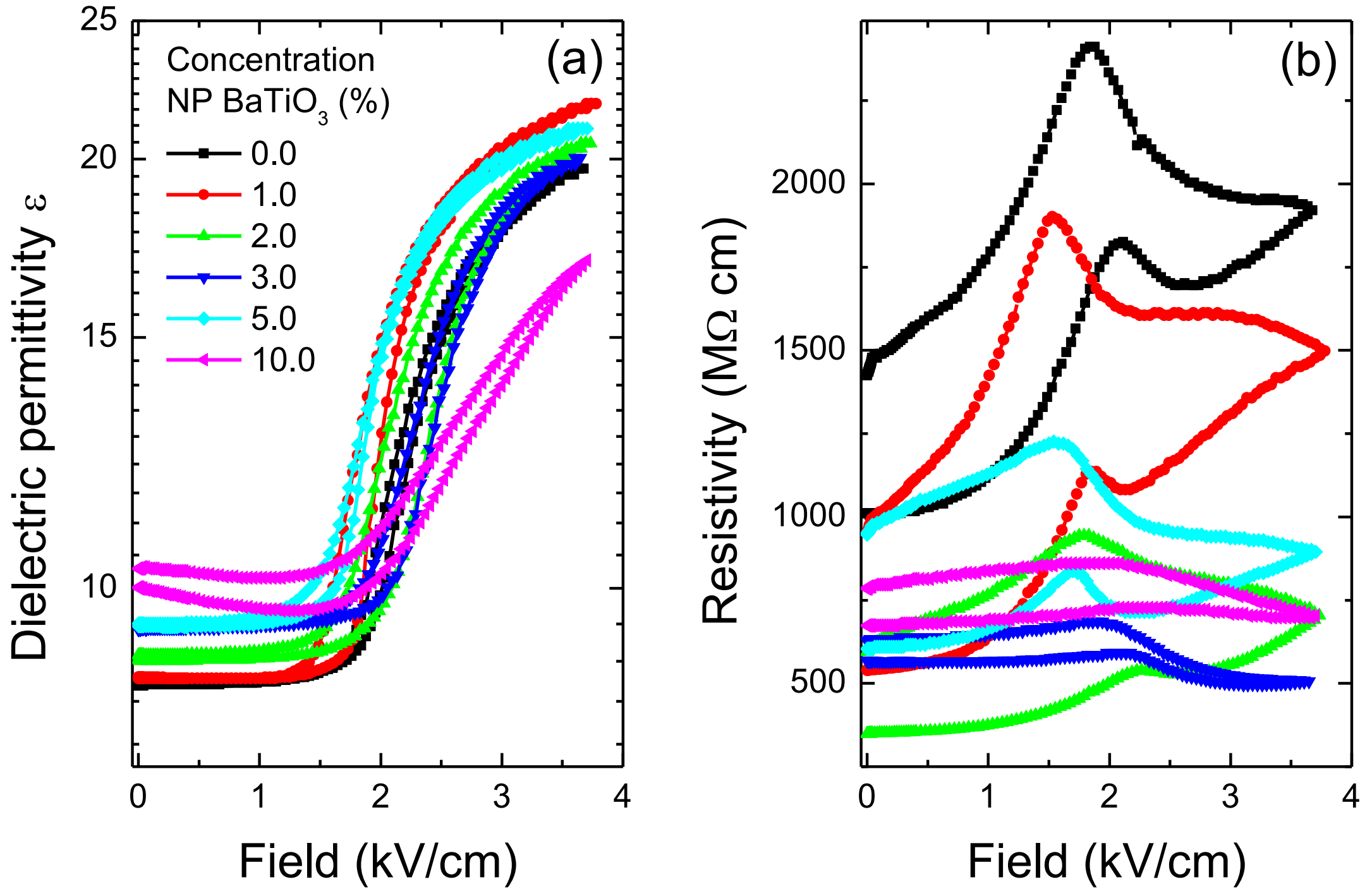

**Figure 4.** Dependences of the effective dielectric permittivity **(a)** and electrical resistivity **(b)** on the electric field strength in the 5CB suspensions with the $BaTiO_3$ nanoparticles measured at 20°C.

**Figure 5** shows the temperature dependences of the effective dielectric permittivity of the 5CB liquid crystal with different concentrations of the $BaTiO_3$ nanoparticles measured under and without external voltage bias of 10 V perpendicular to the cell plane. It demonstrates the effect of nanoparticles on the nematic-isotropic phase transition and on the dielectric anisotropy of the system. The lower curves correspond to the perpendicular permittivity, $\varepsilon_{\perp}$, with the planar orientation of the director, and the upper curves correspond to the parallel permittivity $\varepsilon_{\|}$, when the LC molecules have homeotropic alignment. The difference between them, $\Delta\varepsilon$, determines the dielectric anisotropy of the liquid crystal [10-12]. At lower temperatures, the system is in the nematic phase and permittivity at both orientations slowly changes with an increase in temperature. In the range of 26 – 32°C, the dielectric permittivity for both orientations sharply converges and coincides and then remains almost constant with further increasing temperature. This corresponds to the nematic-isotropic phase transition. Above the transition temperature, dielectric anisotropy is absent, and the system becomes dielectrically isotropic. The addition of even small percentage of nanoparticles (e.g., 0.5 wt.%) changes significantly the phase transition temperature.

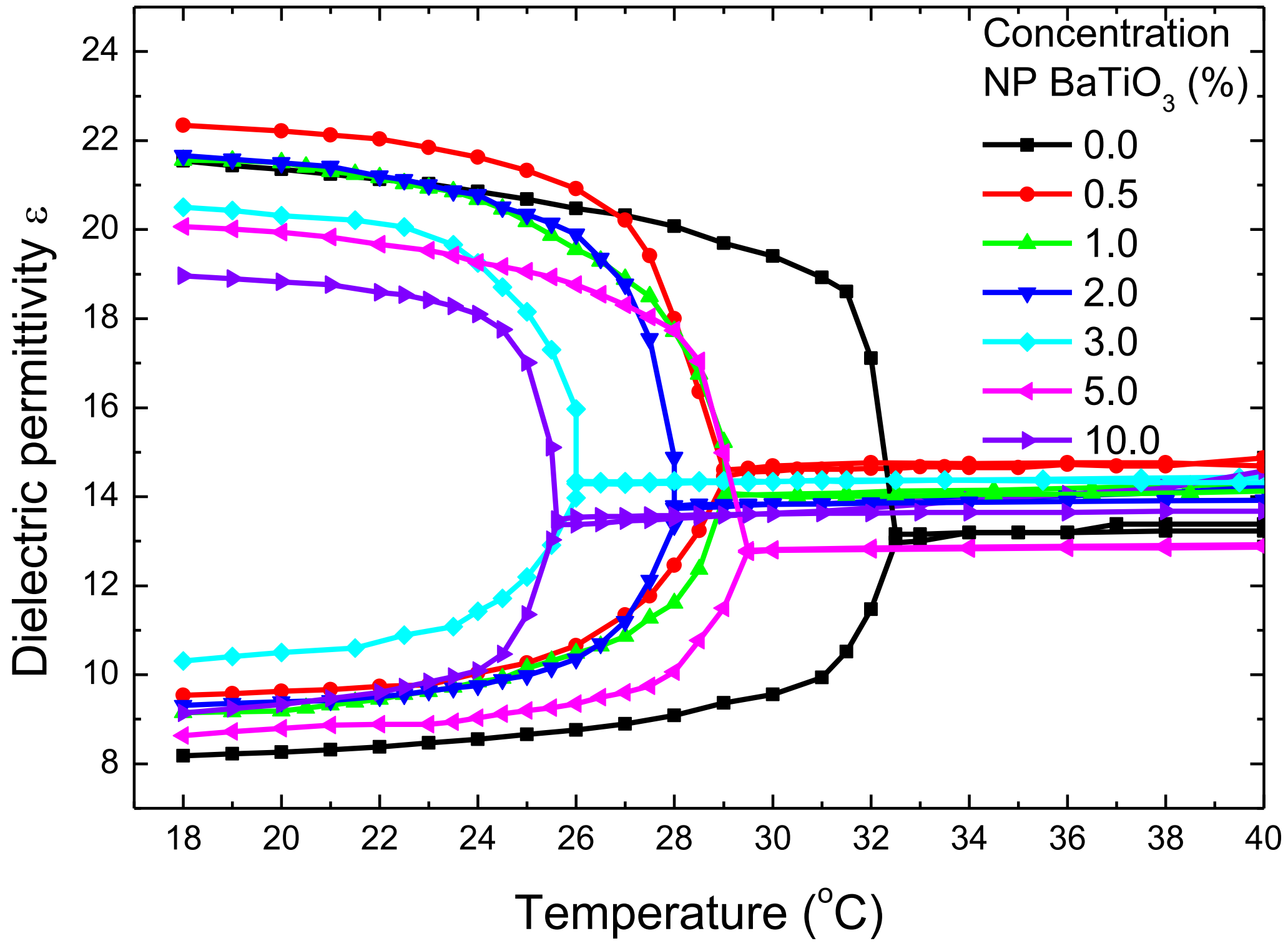

**Figure 5.** Temperature dependences of the effective dielectric permittivity of suspensions with different content of $BaTiO_3$ nanoparticles.

**Figures 6(a)-(c)** present a summary of the results of the study of the influence of the concentration of $BaTiO_3$ nanoparticles on the nematic-isotropic phase transition temperature, anisotropy of the dielectric permittivity, and the dielectric permittivity in the isotropic phase.

**Figure 6(a)** shows the dependence of the nematic-isotropic phase transition temperature on the concentration of $BaTiO_3$ nanoparticles. The phase transition temperature decreases compared to the pure liquid crystal, that means that the nanoparticles destabilize the nematic phase.

**Figure 6(b)** shows the dependence of the dielectric anisotropy $\Delta\varepsilon = \varepsilon_{||} - \varepsilon_{\perp}$ on the concentration of nanoparticles. The magnitude of anisotropy monotonously and significantly decreases with increasing barium titanate nanoparticles content, indicating a stable violation of the orientational order in the system.

**Figure 6(c)** shows the dependence of the effective dielectric permittivity in the isotropic phase. Effective dielectric permittivity increases for small concentrations and begins to decrease at larger concentrations, namely at 5% and 10%.

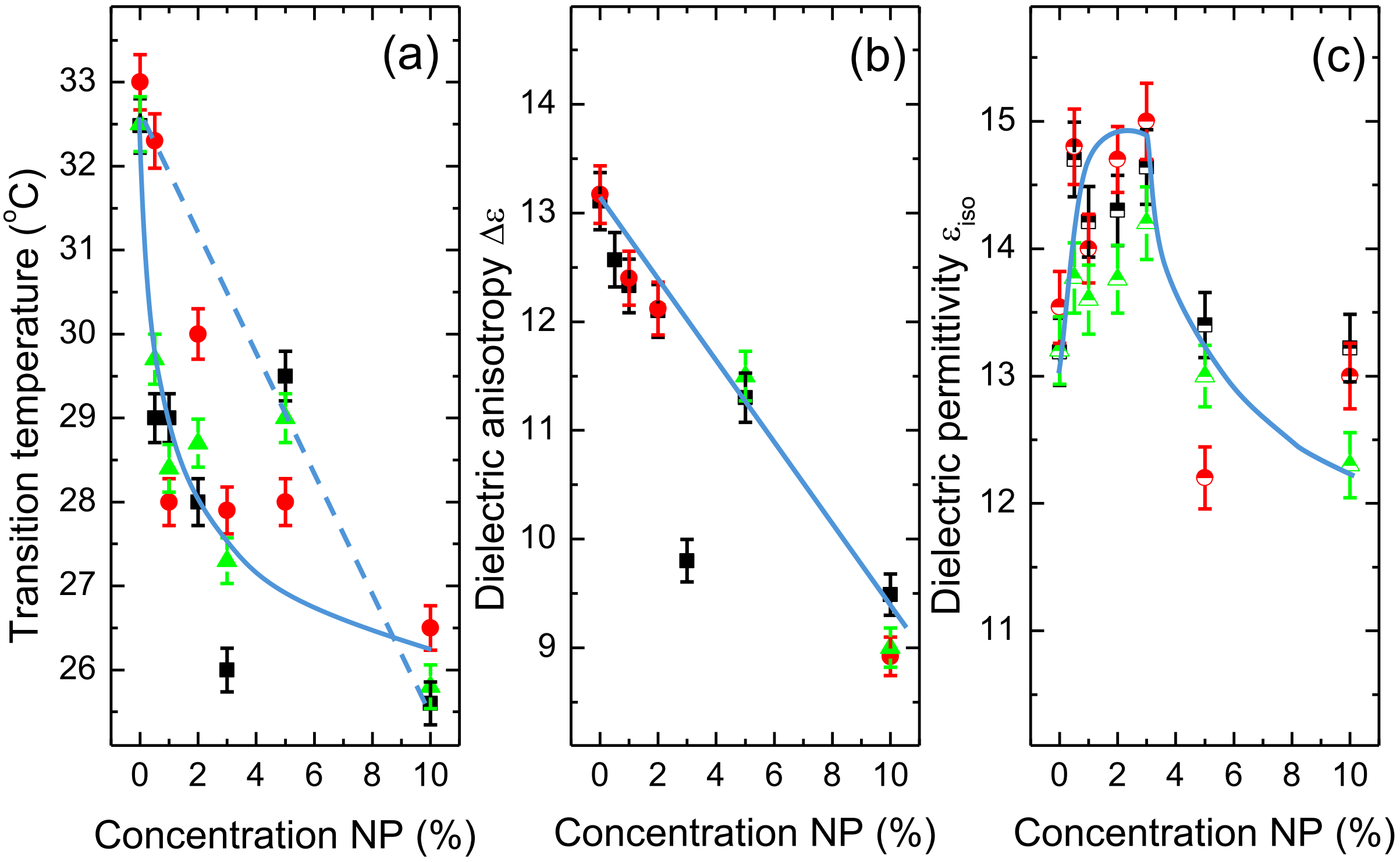

**Figure 6.** Dependence of the nematic-isotropic phase transition temperature **(a)**, dielectric anisotropy **(b)**, and dielectric permittivity in the isotropic phase **(c)** on the concentration of the $BaTiO_3$ nanoparticles. Black, red and green symbols with error bars correspond to the different samples with the same concentration of nanoparticles. Blue solid curves are theoretical fittings,

which details are described in section 4. The dashed blue line in part (a) is added for comparison with the nonlinear dependence shown by the solid blue curve.

## *4. Theoretical interpretation*

Peculiarities of the effective dielectric permittivity shown in **Fig. 6** can be analyzed within the effective media approximation (EMA) [25] considering the interplay of dipolar polarization, possible interplay of the internal barrier layer capacitance (IBLC) and surface barrier layer capacitance (SBLC) effects [26], which are various manifestations of the Maxwell-Wagner (MW) effect [25]. EMA yields an algebraic equation for determining the complex dielectric permittivity $\varepsilon^*_{eff}$ of the effective medium [27]. The real part of $\varepsilon^*_{eff}$ corresponds to the dielectric (polarization) response, and the imaginary part of $\varepsilon^*_{eff}$ accounts for dielectric losses due to finite ionic conductivity, which give rise to MW-type and SBLC effects.

The EMA equation can also be applied to a system of oriented ellipsoidal nanoparticles made of an anisotropic dielectric in an external medium with anisotropic permittivity (see page 200, equation 72 from Ref. [25]):

$$(1-\mu)\frac{\varepsilon^*_{eff}-\varepsilon^*_{LC}}{(1-\eta_{NP})\varepsilon^*_{eff}+\eta_{NP}\,\varepsilon^*_{LC}}+\mu\frac{\varepsilon^*_{eff}-\varepsilon^*_{NP}}{(1-\eta_{NP})\varepsilon^*_{eff}+\eta_{NP}\,\varepsilon^*_{NP}}=0. \qquad (1a)$$

Here $\varepsilon^*_{LC}$ and $\varepsilon^*_{NP}$ as the dielectric permittivity values along the applied field for liquid crystal and nanoparticles, respectively; $\eta_{NP}$ is the depolarization field factor of the nanoparticles in the same direction, at that $0 \le \eta_{NP} \le 1$. The values $\mu$ and $1-\mu$ are relative volume fractions of the nanoparticles "$NP$" and liquid crystal "$LC$", respectively. The solution of quadratic Eq.(1a) is the following [28]:

$$\varepsilon^{\pm}_{eff}=\frac{(1-\mu)\varepsilon^*_{LC}+\mu\varepsilon^*_{NP}-\eta_{NP}(\varepsilon^*_{LC}+\varepsilon^*_{NP})}{2(1-\eta_{NP})}\pm\frac{\sqrt{\left[(1-\mu)\varepsilon^*_{LC}+\mu\varepsilon^*_{NP}-\eta_{NP}(\varepsilon^*_{LC}+\varepsilon^*_{NP})\right]^2+4\varepsilon^*_{LC}\varepsilon^*_{NP}(1-\eta_{NP})\eta_{NP}}}{2(1-\eta_{NP})}. \qquad (1b)$$

The sign "+" corresponds to higher positive value of effective permittivity that corresponds to the thermodynamic equilibrium at real $\varepsilon^*_{LC} > 0$ and $\varepsilon^*_{NP} > 0$. In what follows, one can analyze both solutions to use those corresponding to the deepest energy minimum. The condition $\mathrm{Re}[\varepsilon^*_{eff}] > 1$ should be always true, while the condition $\mathrm{Re}[\varepsilon^*_{NP}] < 0$, that corresponds to a negative capacitance (NC) state of the ferroelectric nanoparticles [29, 30], may have a physical sense for a definite relation between $\mu$, $\varepsilon^*_{LC}$ and $\varepsilon^*_{NP}$.

The nanoparticles are assumed to be monodisperse and uniformly polarized for the validity of Eq. (1b). In the case of randomly oriented/polarized nanoparticles, averaging over the directions of the axes of elliptical nanoparticles is necessary, which yields the following equation (see Eq. (78) on page 202 from Ref. [25]):

$$\sum_{i=1}^{3}\left[\frac{(1-\mu)\eta_i\left(\varepsilon_{eff}^{*}-\varepsilon_{LC_i}^{*}\right)}{(1-\eta_i)\varepsilon_{eff}^{*}+\eta_i\,\varepsilon_{LC_i}^{*}}+\frac{\mu\,\eta_i\left(\varepsilon_{eff}^{*}-\varepsilon_{NP_i}^{*}\right)}{(1-\eta_i)\varepsilon_{eff}^{*}+\eta_i\,\varepsilon_{NP_i}^{*}}\right]=0. \quad (2)$$

Here the summation is performed with index "$i$" denoting different directions $x$, $y$ and $z$; $\varepsilon_{LC_i}^{*}$ and $\varepsilon_{NP_i}^{*}$ are the components of dielectric permittivity tensor along the direction "$i$" for the liquid crystal and nanoparticles respectively, $\eta_i$ are three depolarization factors of the ellipsoid. The value of $\eta_i$ is determined by the shape of nanoparticles and polarization orientation, at that $\sum_{i=1}^{3}\eta_i \equiv 1$. Analytical expressions for $\eta_i$ exist for uniformly polarized nanoparticles of ellipsoidal shape including nanowires, nanospheres, and nanodisks [31]. Screening charges, which are accumulated at the nanoparticle surface, weaken the depolarization field inside the particle, and corresponding decrease of the "effective" depolarization factor $\eta_{eff}$ can be roughly estimated as $\frac{\eta_i}{1+(D/\lambda_{eff})}$, where $D$ is the nanoparticle size in the direction of the spontaneous polarization and $\lambda_{eff}$ is the effective screening length. In particular, $\eta_{eff} < 0.1\eta_i$ for $\lambda_{eff} < 1$ nm and $D$ >10 nm (see **Fig. 7(a)**).

Note that $\eta_i = \frac{1}{3}$ for the spatially isolated and weakly screened ($\lambda_{eff} \gg 1$ nm) spherical nanoparticles, and $\eta_i \to 0$ in the case of their perfect screening by e.g., free ions, hydroxyl groups and/or electrons. In these two cases the solution (1b) for $\varepsilon_{eff}^{\pm}$ has a relatively simple form:

$$\varepsilon_{eff}^{\pm}=\frac{(2-3\mu)\varepsilon_{LC}^{*}+(3\mu-1)\varepsilon_{NP}^{*}}{4}\pm\frac{\sqrt{\left[(2-3\mu)\varepsilon_{LC}^{*}+(3\mu-1)\varepsilon_{NP}^{*}\right]^{2}+8\varepsilon_{LC}^{*}\varepsilon_{NP}^{*}}}{4}, \quad \eta_i=\frac{1}{3}. \quad (3a)$$

$$\varepsilon_{eff}^{*}=(1-\mu)\varepsilon_{LC}^{*}+\mu\varepsilon_{NP}^{*}, \qquad \eta_i \to 0, \quad (3b)$$

Note that polarized and poorly screened nanoparticles can rotate in the external field to align their dipole moments along the direction of external field, and the rotation may lead to the local perturbation of the director of the liquid crystal. Such perturbations, as well as mechanical and/or electric disorder of the director, decrease the nematic-isotropic phase transition temperature. The overall decrease of nematic-isotropic phase transition temperature $T_{tr}$ with increase in $\mu$, shown in **Fig. 6(a)**, is strongly nonlinear (compare solid and dashed blue curves) with a tendency to saturate at the largest $\mu$. Such dependence can be fitted by a hyperbola, namely $T_{tr}=\frac{T_0}{1+(\mu/\mu_0)}$, where $T_0$ and $\mu_0$ are fitting parameters (see solid curve in **Fig. 6(a)**).

Without external electric field, the averaging over equiprobable directions of the ferroelectric nanoparticles spontaneous polarization does not induce any anisotropy, the dielectric anisotropy of the suspension is determined by the anisotropy of the liquid crystal only. Therefore, a monotonic decrease of anisotropy $\Delta\varepsilon_{eff}^{*}$ with increase in $\mu$, shown in **Fig. 6(b)**, can be estimated from Eqs. (3) as $\Delta\varepsilon_{eff}^{*} \cong \frac{(2-3\mu)}{4}\Delta\varepsilon$ for $\eta_i=\frac{1}{3}$ and $\Delta\varepsilon_{eff}^{*} \cong (1-\mu)\Delta\varepsilon$ for $\eta_i \to 0$, where $\Delta\varepsilon$ is the anisotropy of pure liquid crystal without nanoparticles. It is seen from the blue line in **Fig. 6(b)**, that the linear decrease of $\Delta\varepsilon_{eff}^{*}$ with increase in $\mu$ agrees with the experimental measurements considering error bars.

To describe the nonmonotonic dependence of the isotropic permittivity $\varepsilon_{iso}$ versus the nanoparticles concentration $\mu$, shown in **Fig. 6(c)**, consideration of competing mechanisms, which determine the dielectric response for different $\mu$, is required. Schematic image of the competing mechanisms of the dielectric response in the suspension 5CB + $BaTiO_3$ nanoparticles is shown in **Fig. 7(a)**. Note that the cross-over from the ion capturing regime by the ferroelectric nanoparticles to the ion release regime by paraelectric nanoparticles may appear due to the disappearance of their spontaneous polarization with an increase in temperature above the temperature of size-induced phase transition. For instance, the temperature increase from 20°C to 30°C should induce the transition to the paraelectric phase in the 15 – 25 nm nanoparticles, but the specific range may depend on the bias voltage due to the polarization relaxation effects.

Relative effective permittivity $\varepsilon_{eff}^{+}$ as a function of the nanoparticles volume fraction $\mu$ and permittivity $\varepsilon_{NP}$ calculated for $\eta_i = \frac{1}{3}$ and $\eta_i = 0$ without considering losses and assuming that $\varepsilon_{LC} = 13$, is shown in **Fig. 7(b)** and **7(c)**, respectively. It is seen from the figures that $\varepsilon_{eff}^{+}$ increases monotonically with increase in $\mu$ for positive $\varepsilon_{NP}$, or decreases monotonically with increase in $\mu$ for negative $\varepsilon_{NP}$. The monotonic increase of $\varepsilon_{eff}^{+}$ with increase in $\mu$ agrees with the experimental results shown in **Fig. 6(c)** for nanoparticles concentration less than 4 mass %. Since $BaTiO_3$ is approximately 6 times heavier than the LC 5CB, its volume fraction $\mu$ is approximately 6 times smaller. The decrease of $\varepsilon_{eff}^{+}$ with increase in $\mu$ is possible in the NC state of the $BaTiO_3$ nanoparticles. Note that the NC state is stable in the vicinity of the paraelectric-ferroelectric phase transition [32], as well as MW-type IBLC and SBLC effects can lead to the NC state of the ferroelectric nanoparticles with an increase in their concentration. Thus, we can conclude that the nonmonotonic dependence $\varepsilon_{iso}(\mu)$, shown in **Fig. 6(c)**, may be explained by the cross-over between the competing mechanisms of the suspension dielectric response.

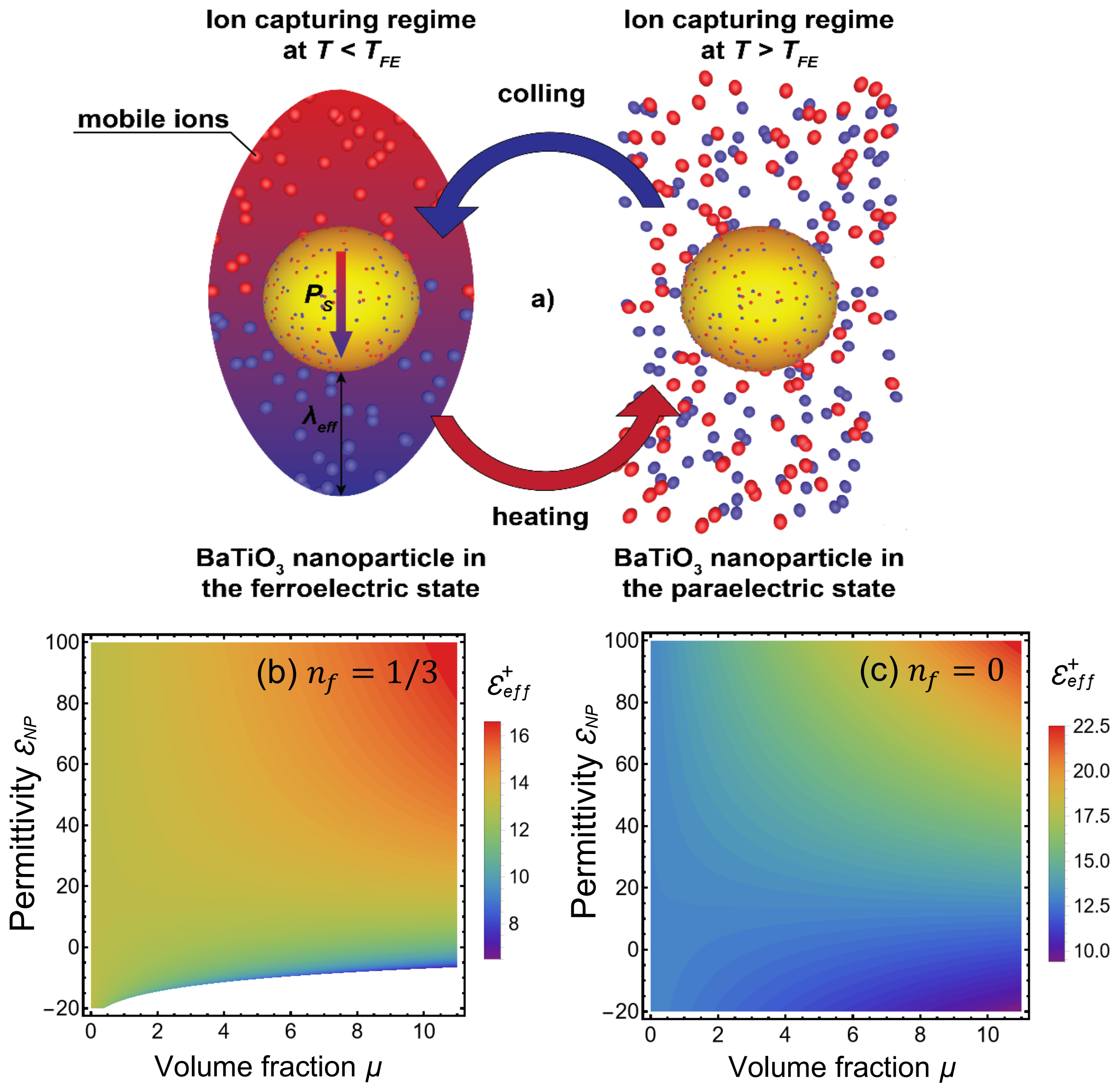

**FIGURE 7**. **(a)** Schematic image of the competing mechanisms of the dielectric response in the suspension 5CB + $BaTiO_3$ nanoparticles. The radial cross-section of the spherical ferroelectric nanoparticle covered with the shell of ionic-electronic screening charges with the effective screening length $\lambda$. The colored arrow shows the direction of the spontaneous polarization $P_S$ inside the nanoparticle. The nanoparticles in the paraelectric phase (i.e., at the temperature $T > T_{FE}$) and in the ferroelectric phase (i.e., at the temperature $T < T_{FE}$) are placed in the liquid crystal 5CB with ions and/or free radicals. Effective dielectric permittivity $\varepsilon_{eff}^{+}$ as a function of the nanoparticles volume fraction $\mu$ and permittivity $\varepsilon_{NP}$ calculated for $\eta_i = \frac{1}{3}$ **(b)** and $\eta_i = 0$ **(c)** without considering losses, and assuming that $\varepsilon_{LC} = 13$.

## 5. *Conclusions*

The obtained results show that ferroelectric $BaTiO_3$ nanoparticles significantly affect the dielectric and conductive properties of the 5CB liquid crystal. At low concentrations, polarization enhancement and dipole-dipole interaction mechanisms dominate. At high concentrations, aggregation of nanoparticles, ionic screening effects and formation of charge transfer channels appear and increase their contribution, which changes the behavior of the colloid and can reduce the effective dielectric response.

**Author's contribution.** The idea of this research belongs to V.V.V. Electrophysical measurements and analysis of experimental results were carried out by J.M.G., O.S.P., V.V.V., D.O.S. and V.N.P. I.A.G. prepared the suspensions, S.E.I. synthesized the nanoparticles. A.N.M. stated the theoretical problem and performed analytical calculations, E.A.E. wrote the codes and performed numerical simulations. Theoretical interpretation of the experimental results is performed by A.N.M. J.M.G., V.V.V. and A.N.M. wrote the manuscript. V.N.P. coordinated the research and improved the text of the article.

**Acknowledgements.** The work of J.M.G., O.S.P., D.O.S., A.N.M. and E.A.E. is funded by the National Science Foundation of Ukraine (grants No. 2023.03/0132 "Multiple degenerated metastable states of spontaneous polarization in nanoferroics: theory, experiment and prospects for digital nanoelectronics" and No. 2023.03/0127 "Silicon-compatible ferroelectric nanocomposites for electronics and sensorics"). The work of V.V.V. and V.M.P. is funded by the target program of the NAS of Ukraine, project No. 5.8/26-P "Energy-saving and environmentally friendly nanoscale ferroics for the development of sensors, nanoelectronics and spintronics". S.E.I. and A.N.M. also acknowledge the NATO Science for Peace and Security Programme under grant SPS G5980 "FRAPCOM" and the Ministry of Education and Science of Ukraine (grant M/14-2026) for sponsoring nanoparticles preparation and characterization.

## References

[1] Y. Reznikov, O. Buchnev, O. Tereshchenko, V. Reshetnyak, A. Glushchenko, and J. West. Ferroelectric nematic suspension. Applied Physics Letters **82**, 1917 (2003); https://doi.org/10.1063/1.1560871

[2] H. Atkuri, G. Cook, D.R. Evans, C.I. Cheon, A. Glushchenko, V. Reshetnyak, Yu. Reznikov, J. West, and K. Zhang. Preparation of ferroelectric nanoparticles for their use in liquid crystalline colloids. Journal of Optics A: Pure and Applied Optics **11**, 024006 (2009); https://doi.org/10.1088/1464-4258/11/2/024006

[3] O. Kurochkin, H. Atkuri, O. Buchnev, A. Glushchenko, O. Grabar, R. Karapinar, V. Reshetnyak, J. West, and Yu. Reznikov. Nano-colloids of $Sn_2P_2S_6$ in nematic liquid crystal pentyl-cyanobiphenyl. Condensed Matter Physics **13**, 33701 (2010); https://doi.org/10.5488/CMP.13.33701

[4] G. Cook, A.V. Glushchenko, V. Reshetnyak, A.T. Griffith, M.A. Saleh, and D.R. Evans. Nanoparticle doped organic-inorganic hybrid photorefractives. Optics Express **16**, 4015 (2008); https://doi.org/10.1364/OE.16.004015

[5] D.R. Evans, S.A. Basun, G. Cook, I.P. Pinkevych, and V.Yu. Reshetnyak. Electric field interactions and aggregation dynamics of ferroelectric nanoparticles in isotropic fluid suspensions. Physical Review B **84**, 174111 (2011); https://doi.org/10.1103/PhysRevB.84.174111

[6] J.M. Gudenko, O.S. Pylypchuk, V.V. Vainberg, I.A. Gvozdovskyy, S.E. Ivanchenko, D.O. Stetsenko, N.V. Morozovsky, V.N. Poroshin, E. A. Eliseev, and A.N. Morozovska. Ferroelectric nanoparticles in liquid crystals: the role of ionic transport at small concentrations of the nanoparticles. Semiconductor Physics, Optoelectronics and Quantum Electronics, **28** (1), 010 (2025); https://doi.org/10.15407/spqeo28.01.010

[7] J. M. Gudenko, O. S. Pylypchuk, V. V. Vainberg, V. N. Poroshin, D. O. Stetsenko, I. A. Gvozdovskyy, O. V. Bereznykov, S. E. Ivanchenko, E. A. Eliseev, A. N. Morozovska. Dynamics of charge states at the surface of a ferroelectric nanoparticle in a liquid crystal. Semiconductor Physics, Optoelectronics and Quantum Electronics, **28** (3), 258 (2025); https://doi.org/10.15407/spqeo28.03.258

[8] Saurabh N. Paul, R. Dhar, R. Verma, S. Sharma & R. Dabrowski. Change in Dielectric and Electro-Optical Properties of a Nematic Material (6CHBT) Due to the Dispersion of $BaTiO_3$ Nanoparticles, Molecular Crystals and Liquid Crystals, **545**, 105/[1329]-111/[1335] (2011); http://dx.doi.org/10.1080/15421406.2011.571961

[9] U.B. Singh, R. Dhar, R. Dabrowski & M.B. Pandey. Enhanced electro-optical properties of a nematic liquid crystals in presence of $BaTiO_3$ nanoparticles. Liquid Crystals, **41** 953 (2014); http://dx.doi.org/10.1080/02678292.2014.894209

[10] A. Rastogi, A. Mishra, F. P. Pandey, R. Manohar, A. S. Parmar, Enhancing physical characteristics of thermotropic nematic liquid crystals by dispersing in various nanoparticles and their potential applications (Review), Emergent Materials **6** 101 (2022); https://doi.org/10.1007/s42247-022-00406-7

[11] A. Drozd-Rzoska, S. Starzonek, S. J. Rzoska, and S. Kralj. Nanoparticle-controlled glassy dynamics in nematogen-based nanocolloids. Physical Review E **99**, 052703 (2019); https://doi.org/10.1103/PhysRevE.99.052703

[12] D. Varshney, J. Prakash, V. P. Singh, K. Yadav, and G. Singh. Probing the impact of bismuth-titanate based nanocomposite on the dielectric and electro-optical features of a nematic liquid crystal material. Journal of Molecular Liquids **347**, 118389 (2022); https://doi.org/10.1016/j.molliq.2021.118389

[13] D. Caruntu, T. Rostamzadeh, T. Costanzo, S.S. Parizi, and G. Caruntu. Solvothermal synthesis and controlled self-assembly of monodisperse titanium-based perovskite colloidal nanocrystals. Nanoscale **7**, 12955 (2015); https://doi.org/10.1039/C5NR00737B

[14] S. Nayak, T.K. Chaki, and D. Khastgir. Spherical ferroelectric $PbZr_{0.52}Ti_{0.48}O_3$ nanoparticles with high permittivity: Switchable dielectric phase transition with temperature. Ceramics International **42**, 14490 (2016); https://doi.org/10.1016/j.ceramint.2016.06.056

[15] A. Lorenz, N. Zimmermann, S. Kumar, D.R. Evans, G. Cook, and H.-S. Kitzerow. Doping the nematic liquid crystal 5CB with milled $BaTiO_3$ nanoparticles. Physical Review E **86**, 051704 (2012); https://doi.org/10.1103/PhysRevE.86.051704

[16] E.S. Beh, S.A. Basun, X. Feng, I.U. Idehenre, D.R. Evans, and M.W. Kanan. Molecular catalysis at polarized interfaces created by ferroelectric $BaTiO_3$. Chemical Science **8**, 2790, (2017); https://doi.org/10.1039/C6SC05032H

[17] I.U. Idehenre, Yu.A. Barnakov, S.A. Basun, and D.R. Evans. Spectroscopic studies of the effects of mechanochemical synthesis on $BaTiO_3$ nanocolloids prepared using high-energy ball-milling. Journal Applied Physics **124**, 165501 (2018); https://doi.org/10.1063/1.5046682

[18] A.N. Morozovska, O.S. Pylypchuk, S. Ivanchenko, E.A. Eliseev, H.V. Shevliakova, L.M. Korolevich, L.P. Yurchenko, O.V. Shyrokov, N.V. Morozovsky, V.N. Poroshin, Z. Kutnjak, and V.V. Vainberg. Electrocaloric response of the dense ferroelectric nanocomposites. Ceramics International **50**, 11743 (2024); https://doi.org/10.1016/j.ceramint.2024.01.079

[19] Yu.A. Barnakov, I.U. Idehenre, S.A. Basun, T.A. Tyson, and D.R. Evans. Uncovering the mystery of ferroelectricity in zero dimensional nanoparticles. Nanoscale Advances **1**, 664 (2019); https://doi.org/10.1039/C8NA00131F

[20] H. Zhang, S. Liu, S. Ghose, B. Ravel, I.U. Idehenre, Y.A. Barnakov, S.A. Basun, D.R. Evans, and T.A. Tyson. Structural Origin of recovered ferroelectricity in $BaTiO_3$ nanoparticles. Physical Review B **108**, 064106 (2023); https://doi.org/10.1103/PhysRevB.108.064106

[21] E.A. Eliseev, A.N. Morozovska, S.V. Kalinin, and D.R. Evans. Strain-induced polarization enhancement in $BaTiO_3$ core-shell nanoparticles. Physical Review B **109**, 014104 (2024); https://doi.org/10.1103/PhysRevB.109.014104

[22] A. N. Morozovska, E. A. Eliseev, O. A. Kovalenko, and D. R. Evans. The Influence of Chemical Strains on the Electrocaloric Response, Polarization Morphology, Tetragonality and Negative Capacitance Effect of Ferroelectric Core-Shell Nanorods and Nanowires. Physical Review

Applied **21**, 054035 (2024); https://doi.org/10.1103/PhysRevApplied.21.054035

[23] O. A. Kovalenko, E. A. Eliseev, Y. O. Zagorodniy, S. D. Škapin, M. M. Kržmanc, L. Demchenko, V. V. Laguta, Z. Kutnjak, D. R. Evans, and A. N. Morozovska. Strain-Gradient and Curvature-Induced Changes in Domain Morphology of $BaTiO_3$ Nanorods: Experimental and Theoretical Studies, Physical Review Materials **10**, 044409 (2026); https://doi.org/10.1103/b332-gcxc

[24] S. V. Kalinin, E. A. Eliseev, and A. N. Morozovska. Adsorption of ions from aqueous solutions by ferroelectric nanoparticles. Physical Review Applied **23**, 014081 (2025); https://doi.org/10.1103/PhysRevApplied.23.014081

[25] G. L. Carr, S. Perkowitz, and D. B. Tanner. Far-infrared properties of inhomogeneous materials. In "Infrared and millimeter waves" Orlando, FL, Academic Press, Inc. **13** (A86-37151 17-33) 171 (1985).

[26] J. Petzelt, D. Nuzhnyy, V. Bovtun, M. Savinov, M. Kempa, I. Rychetsky, Broadband dielectric and conductivity spectroscopy of inhomogeneous and composite conductors. Phys. Stat. Sol. A **210**, 2259 (2013); https://doi.org/10.1002/pssa.201329288

[27] O. S. Pylypchuk, V. V. Vainberg, V. N. Poroshin, O. V. Leshchenko, V. N. Pavlikov, I. V. Kondakova, S. E. Ivanchenko, L. P. Yurchenko, L. Demchenko, A. O. Diachenko, M. V. Karpets, M. P. Trubitsyn, E. A. Eliseev, and A. N. Morozovska. A colossal dielectric response of HfxZr1-xO2 nanoparticles. Physical Review Materials **9**, 114412 (2025); https://doi.org/10.1103/y2pb-5g5w

[28] A. N. Morozovska and M. V. Strikha. Influence of the negative capacitance state of the layer with ferroelectric nanoparticles on the FET subthreshold swing. J. Phys.: Conf. Ser. **3263**, 012029 (2026); https://doi.org/10.1088/1742-6596/3263/1/012029

[29] O. V. Bereznykov, O. S. Pylypchuk, V. I. Styopkin, S. E. Ivanchenko, Denis O. Stetsenko, E. A. Eliseev, Z. Kutnjak, V. N. Poroshin, A. N. Morozovska, and N. V. Morozovsky. Interfacial Effects and Negative Capacitance State in P(VDF-TrFE) Films with $BaTiO_3$ Nanoparticles. Composites Interfaces, **33** (3), 621 (2026); https://doi.org/10.1080/09276440.2025.2538943

[30] A. N. Morozovska, O. V. Bereznykov, M. V. Strikha, O. S. Pylypchuk, Z. Kutnjak, E. A. Eliseev, and D. R. Evans. Domain Morphology, Electrocaloric Response, and Negative Capacitance States of Ferroelectric Nanowires Array. Composite Interfaces **33** 1943 (2025); https://doi.org/10.1080/09276440.2025.2605389

[31] L.D. Landau, E.M. Lifshitz, L. P. Pitaevskii. Electrodynamics of Continuous Media, (Second Edition, Butterworth-Heinemann, Oxford, 1984).

[32] A. N. Morozovska, S. Cherifi-Hertel, E. A. Eliseev, V. V. Khist, R. Hertel, and D. R. Evans. The Role of Flexoelectric Coupling and Chemical Strains in the Emergence of Polar Chiral Nano-Structures. Journal of Applied Physics, **138**, 030701 (2025); https://doi.org/10.1063/5.0277483